
\input phyzzx

\catcode`\@=11
\def\eqaligntwo#1{\null\,\vcenter{\openup\jot\m@th
\ialign{\strut\hfil
$\displaystyle{##}$&$\displaystyle{{}##}$&$\displaystyle{{}##}$\hfil
\crcr#1\crcr}}\,}
\catcode`\@=12

\def\NP{{\it Nucl. Phys.\ }}

\def\PL{{\it Phys. Lett.\ }}

\def\PRL{{\it Phys. Rev. Lett.\ }}

\def\IJMP{{\it Int. Jour. Mod. Phys.\ }}
\def\Mod{{\it Mod. Phys. Lett.\ }}

\REF\pol{A. M. Polyakov, \Mod {\bf A6} (1991) 635.}
\REF\minic{D. Minic and Z. Yang, Texas preprint UTTG-23-91.}
\REF\gkn{D.J. Gross, I.R. Klebanov and M.J. Newman, \NP {\bf B350}
(1991) 621.}
\REF\kdf{
 P. Di Francesco and D. Kutasov, \PL {\bf 261B} (1991) 385.}
\REF\kd{P. Di Francesco and D. Kutasov, Princeton preprint
PUPT-1276.}
\REF\Goul{M. Goulian and M. Li, \PRL {\bf 66} (1991) 2051.}
\REF\gupta{A. Gupta, S. P. Trivedi, and M. B. Wise,
\NP {\bf B340} (1990) 475.}
\REF\BK{M. Bershadsky and I. R. Klebanov, \PRL {\bf 65} (1990) 3088;
N. Sakai and Y. Tanii, \IJMP {\bf A6} (1991) 2743.}
\REF\Gk{D.J. Gross and I.R. Klebanov, \NP {\bf B359} (1991) 3.}
\REF\igorsasha{I.R. Klebanov and A.M. Polyakov,
\Mod {\bf A6} (1991) 3273.}
\REF\mats{Y. Matsumura, N. Sakai and Y. Tanii, Tokyo preprint
TIT/HEP-186, STUPP-92-124.}
\REF\moore{G. Moore, Rutgers preprint RU-91-12 (1991);
G. Moore, R. Plesser, and S. Ramgoolam, Yale
preprint YCTP-P35-91; G. Moore and R. Plesser, Yale preprint
YCTP-P7-92.}
\REF\das{J. Polchinski, \NP {\bf B346 } (1990) 253;
S. R. Das and A. Jevicki, \Mod {\bf A5} (1990) 1639.}
\REF\GK{D.~J.~Gross and I.~R.~Klebanov, \NP {\bf B344} (1990) 475.}
\REF\joe{J. Polchinski, \NP {\bf B346} (1990) 253.}
\REF\poly{A. M. Polyakov, Princeton preprint PUPT-1289}
\REF\igor{I.R. Klebanov, Princeton preprint PUPT-1271.}

\def \a{ \alpha}
\def \b{\beta}
\def \d{\partial}

\def \e{\epsilon}
\def \g{\Gamma}

\def \k{{\bf k}}
\def \E{\varepsilon}
\def \t{ \tau }
\def \f{{\cal F}}
\def \half{ {1\over 2}}
\def \im{ {\rm \;Im \;}}

\def \tt{ \t_2}

\date={April 1992}
\Pubnum={PUPT-1316}
\titlepage

\title{ Unitarity Relations in c=1 Liouville Theory}

\author{David A. Lowe}
\JHL

\abstract

We consider the S-matrix of $c=1$ Liouville theory with vanishing
cosmological constant. We examine some of the constraints imposed by
unitarity. These completely determine $(N,2)$ amplitudes at tree level
in terms of the $(N,1)$ amplitudes when the \lq\lq plus'' tachyon momenta
take generic values.
A surprising feature of the matrix model results is the lack of
particle creation branch cuts in the higher genus amplitudes.
In fact, we show the naive field theory limit of Liouville theory would
predict such branch cuts.
However, unitarity in the full string theory ensures
that such cuts do not appear in genus one $(N,1)$ amplitudes.
We conclude with some comments about the genus one $(N,2)$ amplitudes.

\endpage
\section{Introduction}

The Liouville model of $c=1$ quantum gravity
with vanishing cosmological constant
describes a sum over
Euclidean random surfaces coupled to a non-compact scalar boson. With
this simple physical interpretation one would expect this to be a
quantum theory described by a unitary S-matrix.
Following Polyakov [\pol] and Minic and Yang [\minic] we consider
the S-matrix defined by the resonant Liouville tachyon
scattering amplitudes. Because the discrete states [\pol,\gkn] appear
as intermediate states at poles
in these amplitudes it is necessary to include these in the Hilbert
space of our theory, and hence in the S-matrix. This kind of S-matrix
resembles what we see in more realistic string theories, i.e.
amplitudes with
poles due to on-shell intermediate states.
We may hope to obtain more direct insight into higher
dimensional strings
by studying this kind of theory.
It should be noted that the S-matrices considered
in matrix model work [\moore,\das,\Gk] do not possess this property.

{}From the Liouville point of view the only closed string
amplitudes that have been
computed are the $(N,1)$ tachyon amplitudes on the sphere [\pol-\Goul]
, the
genus one partition function [\gupta,\BK], and the three point
couplings of discrete states on the sphere [\igorsasha, \mats].
In this paper we use the unitarity
constraints to extract information about some other
correlation functions.  We find that unitarity completely
determines the $(N,2)$ tachyon amplitudes on the sphere in terms
of the $(N,1)$ amplitudes when the \lq\lq plus'' tachyon momenta
take generic values.
We then consider the integral expression for genus one Liouville amplitudes.
To leading order in the tachyon potential, the tachyon part
of the field theory limit of these amplitudes is shown to correspond
to a massless $\varphi^3$ theory with a time dependent coupling.
This would lead one to expect log branch cuts in the momenta. We show
unitarity conditions in the full string theory imply these cuts do not
appear for the case of $(N,1)$ amplitudes at genus one. We conclude
with some comments about the genus one $(N,2)$ amplitudes.

\section{The S-Matrix and Unitarity}

Let us begin by reviewing the form of the tree-level tachyon amplitudes in
resonant Liouville theory coupled to $c=1$ matter.
Tachyon vertex operators take the form
$$
T(p) = \exp(ip X + (-1+\E) \phi),
$$
where $X$ is the matter field and $\phi$ the Liouville field.
For positive (negative) chirality tachyons $\E=p$ ($\E=-p$).
For $(N-1,1)$ amplitudes, i.e. $N-1$ positive
chirality particles $T^+$ and one negative chirality particle $T^-$
the answer is [\pol,\kdf,\kd,\poly]
$$
\VEV{T^+(p_1)\cdots T^+(p_{N-1}) T^-(p_N)} =
{{\pi^{N-3} } \over {(N-3)!}} z(p_1) \cdots
z(p_{N-1}),
\eqn\npoint$$
where $z(p) = {{\g(1-2p)} / {\g(2p)}}$.
The derivation of this formula does not require us to restrict to
$p_i>0$, so we take $p_i \in (-\infty , \infty)$ which
corresponds to \lq \lq wrongly" dressed
tachyons for $p_i<0$.
A similar formula holds for
$(1,N-1)$ amplitudes with $z(p) \to z(-p)$.
If we take the limit $p_i \to {n / 2}$
with $n$ a positive integer, the amplitude approaches a
single pole coming from an on-shell intermediate state.
Adopting an $i\e$ prescription for the string propagator amounts
to the replacement
$$
{1\over {L_0+{\bar L}_0-2}} \to {1\over {L_0+{\bar L}_0-2+i\e}}.
\eqn\prop
$$
So \npoint\ will develop an imaginary part as a pole is approached
corresponding to a shift $z(p) \to z(p+i\e)$.

We now use the amplitudes of resonant Liouville theory to
define a S-matrix to describe the scattering of tachyons
and the discrete states.
Here we are referring to the discrete states of the form
$V^{\pm}_{j,m}=P_{j,m}(X)P_{j,m}({\bar X}) \exp(imX+(-1\pm j)\phi)$
in which $P_{j,m}(X)$ is a polynomial
in the derivatives of $X$, and $j$ and $m$ are
$SU(2)$ quantum numbers as discussed in [\igorsasha].

In the $\mu =0$ theory we obtain a delta
function expressing conservation of energy by Wick rotating the
$\phi$ zero mode, $\phi_0 \to i \phi_0$ and integrating along
the real axis. This leads to a physical picture of particle
scattering since now the tachyon wavefunctions look like
$\exp(i \E \phi_0 + i p X_0) $.
$X_0$ is interpreted as the space direction and $\phi_0$ as time.
+ tachyons with $p>0$ then correspond to right moving in states
while those with $p<0$ should be interpreted as left moving
out states. Similarly $-$ tachyons with $p<0$ are left moving
in states, and $p>0$ $-$ tachyons are right moving out states.

We choose the tachyon states to be normalized as
$$
\VEV{T(p^\prime) | T(p) } = 4\pi |p| \delta (p-p^\prime ),
\eqn\eq$$
and the tachyon propagator is $i / (\E^2 - p^2+ i\e)$.
Decomposing the S-matrix as
$$
S=1+iR,
\eqn\eq$$
we define the connected part of $R$ corresponding to tachyon scattering
as
$$
\VEV{ \prod_i T(p_i) \biggl| R_c \biggr|\prod_j T( p_j) }
= \pi \VEV{ \prod_i T(p_i) \prod_j T(p_j) },
\eqn\eq$$
where $\VEV{ \prod T(p_i) \prod T(p_j) }$ is the
resonant Liouville scattering
amplitude, and we have suppressed the delta function expressing
conservation of energy and momentum. The factor of $\pi$ is fixed by
demanding consistency of the unitarity relations with the
normalizations implicit in \npoint\ .

Unitarity is an important physical constraint for the
consistency of our theory. Imposing $S S^{\dag} = 1$ gives
$$
R-R^{\dag} = i R R^{\dag} = i R^{\dag} R,
\eqn\eq$$
or equivalently
$$
2 \im \VEV{ \prod_i T(p_i) \biggl| R \biggr| \prod_j T(p_j) } =
\sum_n \VEV{\prod_i T(p_i) \biggl| R\biggr| n}
\VEV{n \biggl| R \biggr| \prod_j T(p_j)},
\eqn\unit$$
where we sum over all positive energy
intermediate states allowed by overall conservation
of
energy and momentum. For tachyon intermediate states this boils
down to replacing the tachyon propagator by
$2 \pi \delta(\E^2-p^2) \theta (\E) $. For discrete states the
propagator is replaced by $2 \pi \delta(\E^2 -p^2-M^2) \theta(\E) {\cal
P}_M$ for
a discrete state of mass $M$. ${\cal P}_M$ is an operator which projects
onto the physical discrete values of $\E$ and $p$ for a state of mass
$M$.

\section{(N,2) Amplitudes on the Sphere}

We now prove that the relation \unit\ is sufficient to determine
all tree-level $(N,2)$ amplitudes when the
+ tachyons take generic values of momenta. Energy and momentum conservation
give
$$
\eqalign{
\sum p_i &= N/2, \cr
\sum k_i &= -N/2. \cr}
\eqn\eq
$$
Here $p_i$ refer to + tachyons and $k_i$ to $-$ tachyons.
It has been argued in [\poly] that these amplitudes should vanish up to
delta function terms. By considering the integral representation
it may be seen that these delta function terms arise due to
on-shell intermediate + tachyons as the $k_i \to -n_i/2$.
Here $n_i$ is a positive integer. There is
no contribution from the discrete states as long as the $p_i$
are kept at generic values.

For a general $(N,2)$ amplitude a typical diagram is shown in Fig 1.
Summing over the permutations of $k_1$ and $k_2$ eliminates
the $\theta(\E)$ factor from the on-shell tachyon propagator.
This leaves us with a sum over the subdivisions $S_1$, $S_2$ of the
set $S$ of + tachyon momenta which satisfy $|S_1|=n$. Let
$q(S)$ denote the sum over the momenta in set $S$.
The unitarity relations \unit\ in the limit $k_1 \to -n/2$ then lead to
$$
\eqalign{
&2 \pi \im \VEV{T^+(p_1) \cdots T^+(p_N) T^-(k_1) T^-(k_2)}  \cr
& =\sum_{{S_1 \sqcup S_2=S}\atop{|S_1|=n}}
2 \pi^3 \VEV{\prod_{S_1} T^+(p_i)  T^+(-q(S_1)-k_1)
T^-(k_1)}
\delta\bigl((n+2k_1)(n-2q(S_1))\bigr)  \cr
& \qquad \qquad \qquad \qquad \qquad \times \VEV{T^+(q(S_1)+k_1)
\prod_{S_2}  T^+(p_j)
 T^-(k_2) } \cr
&= \sum_{{S_1 \sqcup S_2=S}\atop{|S_1|=n}}
 {{2 \pi^{N+1}}\over {(n-1)!(N-n-1)!}}
 z(p_1)\cdots z(p_N)
z(q(S_1)+k_1) z(-q(S_1)-k_1)  \cr
& \qquad\qquad\qquad\qquad\qquad \times \delta \bigl(
(2q(S_1)-n)(n+2k_1)\bigr) \cr
&= \sum_{{S_1 \sqcup S_2=S}\atop{|S_1|=n}}
 -{{4 \pi^{N+1}}\over {(n-1)!(N-n-1)!}}
 z(p_1)\cdots z(p_N) \; | q(S_1)
+ k_1| \; \delta(n+2k_1), \cr }
\eqn\twominus
$$
where we used the fact
that $z(q)z(-q)= -4q^2$.
In fact we can see that the real part of this amplitude should vanish.
Factorizing the amplitude in the limit $k_1 \to -n/2$
and keeping only the real part we obtain
$$
\eqalign{
& \sum_{{S_1 \sqcup S_2=S}\atop{|S_1|=n}}
\VEV{\prod_{S_1} T^+(p_i)  T^+(-q(S_1)-k_1)
T^-(k_1)}
{1\over {(n+2k_1)(n-2q(S_1))}}  \cr
& \qquad \qquad \qquad \qquad \qquad \times \VEV{T^+(q(S_1)+k_1)
\prod_{S_2}  T^+(p_j)
 T^-(k_2) } \cr
&= \sum_{{S_1 \sqcup S_2=S}\atop{|S_1|=n}}
z(p_1)\cdots z(p_N)
{{\pi^{N-2}}\over {(n-1)! (N-n-1)!}}
 {{z(q(S_1)+k_1) z(-q(S_1)-k_1)} \over
 {(n+2k_1)(n-2q(S_1))}}
\cr
&=z(p_1)\cdots z(p_N)
{{2 \pi^{N-2}}\over {(n-1)! (N-n-1)!}}
{{N!} \over {n! (N-n)!}}
{{k_1+ {n\over N} \sum p_i}\over {n+2k_1}} .  \cr}
\eqn\realpart
$$
Recalling $\sum p_i= N/2$ we see the real part of the
residue of the pole vanishes as
$k_1 \to -n/2$, so the imaginary part \twominus\ is the full amplitude.

Let us compare this to the matrix model predictions for non-resonant
amplitudes at tree-level [\kdf, \kd,\moore,\das,\igor]
$$
\eqalign{
 & \VEV{T^+(p_1)\cdots T^+(p_N) T^-(k_1) \cdots T^-(k_M) } = \cr
  &\quad \quad  \mu^s  (-\pi)^{N+M-3}
z(p_1)\cdots
z(p_{N})  z(-k_1) \cdots z(-k_M)
P(\{p_i\},\{k_j\}), \cr }
\eqn\nonres$$
where $P(\{p_i\},\{k_j\})$ is a polynomial in the momenta
and $s=\sum p_i - \sum k_j +2-N-M$.
First we recall how the $\phi$ zero mode integrations are related.
In the case of non-resonant amplitudes we have [\Goul]
$$
\int_{-\infty}^{\infty} d\phi_0 \exp(s\phi_0 - {\Delta} e^{-\phi_0} )
= {\Delta}^s \g(-s) ,
\eqn\eq
$$
if we take $\Delta>0$. Here $\Delta$ is the
bare cosmological constant, related to the renormalized cosmological
constant appearing in \nonres\ by $\Delta = \mu \log(\mu)$ as
discussed in [\igor]. Alternatively we may take
$\phi \exp(-\phi)$ to be the cosmological constant operator.
We see therefore that the energy conserving delta function $\delta(s)$
of the resonant amplitude is replaced by a pole as $s\to 0$.
The rule for comparing the resonant amplitudes and the $s\to 0$ limit
of the non-resonant amplitudes is then: extract the residue of the
$-1/s$ pole and insert the appropriate $i\e$ terms coming from the
shifted string propagator.

For the
$(N,1)$ case this is straightforward.
Kinematics fixes $k_1=-(N-1+s)/2$
so the $-1/s$ pole comes from the $z(-k_1)$ factor in \nonres\ .
For the general $(N,M)$ amplitude with $N>1$ and $M>1$
there is no $-1/s$ pole unless
we impose an extra constraint on the momenta which will give rise
to nonanalytic terms. Therefore these amplitudes are usually
analytically continued to zero.

Let us analyze the nonanalytic contributions for
$(N,2)$ amplitudes when the $p_i$ take generic values.
In the limit $s\to 0$ and $k_1 \to -n/2$ (with $n$ a positive integer)
\nonres\ will lie on a double pole
$$
\eqalign{
&\VEV{T^+(p_1)\cdots T^+(p_N) T^-(k_1) T^-(k_2) } =
 \mu^s \pi^{N-1}  z(p_1)\cdots z(p_N) P(\{p_i\},\{k_j\} )  \cr
& \times {1\over {((n-1)! (N-n-1)!)^2}} \biggl( {1\over {2k_1+n}} +
{1\over {-2k_1-n-s}} \biggr) {1\over {-s}} .\cr}
\eqn\eq
$$
Inserting the appropriate
$i\e$ terms in the factors multiplying $-1/s$ we obtain for
the resonant amplitude
$$
-\pi^{N-1}  z(p_1)\cdots z(p_N) P(\{p_i\} , \{k_j\} ) \;
{1\over {((n-1)! (N-n-1)!)^2}}
\; 2\pi i \; \delta(n+2k_1).
\eqn\eq
$$
Therefore we see that the real part of the amplitude vanishes and
the imaginary part should agree with that computed using
the unitarity relations \twominus\ . This has been checked
for the four and five point functions. For example as $k_1\to -1/2$
the resonant
$(3,2)$ amplitude in any kinematic configuration can be written
$$
\eqalign{
&\VEV{T^+(p_1) T^+(p_2) T^+(p_3) T^-(k_1) T^-(k_2) } = \cr
&-2 \pi^3 i \; z(p_1) z(p_2) z(p_3) \bigl( |p_1+k_1| + |p_2+k_1| +
|p_3+k_1| \bigr) \delta(1+2k_1), \cr}
\eqn\fivept
$$
in agreement with \twominus\ .

\section{Tachyon Effective Field Theory}

Now we investigate the tachyon effective action that
follows from the field theory limit of the one loop correlators.
The leading term in the tachyon potential will turn out to be a simple
cubic. This would lead us to expect log branch cuts in the momenta at one
loop in obvious disagreement with matrix model results. We will
show however that the unitarity relations imply these cuts
are absent in the full string theory indicating a very
subtle interplay between the tachyon field and the discrete states in
this kind of effective theory.

Consider then the formula for the genus one amplitudes in Liouville theory
of tachyon operators of the form $\exp(ipX+(-1+\E) \phi)$
$$
\int_{\f} {{d^2 \t} \over {\tt^2}} \prod_{i=1}^N \int d^2 z_i
\prod_{1 \leq j < i} \chi(z_i -z_j \mid \t )^{-4 \k_i
\cdot \k_j} , \eqn\go
$$
with
$$
\chi (z \mid \t) = \exp \biggl( { {- \pi (\im z )^2 } \over \tt } \biggr)
\biggl\vert
{{\theta_1 (z \mid \t) } \over { \theta_1^\prime ( 0 \mid \t) }} \biggr\vert
,\eqn\ch$$
and
$$
\k=(-1+\E,k), \qquad \qquad \qquad
\k_i \cdot \k_j = (-1+\E_i)(-1+ \E_j) - k_i k_j,
\eqn\do$$
where $\tt= \im \t$ and ${\cal F}$ is the fundamental region
of the modular group $\tt>0$, $|\t|>1$ and $-1/2\leq {\rm \; Re\;} \t < 1/2$.
We may take the field theory limit by sending
the lower cutoff to zero
and extracting the leading asymptote of the integrand as $\tt \to \infty$.
This should correspond to keeping only the purely tachyonic contribution
to \go\ .
For $z = x + \t y$, and $y $ fixed as $\tt \to \infty$,
$$
\chi (z \mid \t) \to \exp ( \pi \tt y(1-y) ) \bigl| 1 - \exp (2 \pi i z)
\bigr|.
\eqn\lim$$
When taking the field theory limit we discard the $ \exp (2 \pi i z) $
term which is suppressed with respect to the leading term
except when $y \to 0$, when \lq \lq contact" terms appear.
This corresponds to only keeping the ring diagram contribution.
Substituting \lim\
into the genus one integral and trivially doing the $\tau_1$ and
$x_i$ integrals we get
$$
\int_0^\infty d \tt \tt^{N-2} \prod_{i=1}^N \int_0^1 d y_i \exp(-4 \pi
\tt \sum_{i<j} |y_i-y_j| (1-|y_i - y_j|) \k_i \cdot \k_j ).
\eqn\eq$$
For this to be well defined we must pick a specific ordering of the
$y_i$, lets say $y_1 < y_2 < \cdots < y_N $, and continue the $k_i$ to a
region where the integral converges.

Now we change variables to the Feynman parameters
$$
\eqalign{
\a_i &= y_{i+1} - y_i , \quad \quad \quad i < N \cr
\a_N &= 1 - y_N + y_1 .\cr }
\eqn\eq$$
The term in the exponential becomes
$$
\sum_{i < j} (y_j - y_i)(1 - (y_j-y_i)) \k_i \cdot \k_j =
\sum_{i < j} \a_i \a_j \b_{i j },
\eqn\eq$$
where
$$
\b_{i j} = \sum_{{{ l > i, \; m > j}\atop{ m>l, \; l \leq j}} }
\k_l \cdot \k_m +
           \sum_{ {{l \leq i, \; m \leq j}\atop {  m>l, \; m > i}} }
\k_l \cdot  \k_m,
\eqn\eq$$
and we finally get
$$
\int_0^\infty d \tt \tt^{N-2} \prod_{i=1}^N \int_0^1 d \a_i
\delta (\sum \a_i -1) \exp( -4 \pi \tt \sum_{l<m} \a_l \a_m \b_{ lm} ),
\eqn\eq$$
which is just the proper time representation for the ring graph of a
massless $\varphi^3$ theory
with a time dependent coupling $g_{st} = g_0 e^{-\phi}$.
Thus to leading order in the tachyon
potential the
effective action for the tachyon field for zero cosmological constant is
$$
S = {1 \over {2g^2_0 }} \int d x
d \phi~ e^{2 \phi} \bigl( (\d_x T)^2 + (\d_{\phi}
T)^2 - T^2 - {1 \over 3} T^3 \bigr),
\eqn\eq$$
which is familiar form from the work of [\joe,\poly].
Naturally there will be higher order corrections to the potential $V(T)$
and interaction terms with the discrete states
but if we only consider the cubic interaction term we expect to see
logarithmic cuts in the momenta at higher genus.
For example evaluating
the three point ring diagram we find a branch cut that goes like
$ \log(1-2p) \theta(1-2p)  /|1-2p| $ as $p\to \half$.
These cuts are not observed in the matrix model and we
would like to understand their absence from the Liouville point of view.

\section{Absence of Branch Cuts at Genus One}

Let us apply the unitarity relations \unit\ to the $(N,1)$
amplitudes at genus one.
Fig 2 shows the type of diagram to be considered.
We take the momenta of the + external
tachyons $p_i$ to be generic. Branch cuts will then appear as finite
contributions to the discontinuity in the
imaginary part of the amplitude.
Note that when some of the $p_i$ lie at discrete values delta function
terms will appear corresponding to poles in the full amplitude, however
this
is the same situation as at tree-level.

Let us consider the different possible combinations of intermediate states.
Two + tachyons are forbidden by kinematics.
We have already seen
from the form of the tree level amplitudes that when $-$ tachyons
appear as intermediate states they will be forced to lie at
discrete values of momenta.
In fact, the situation when both the intermediate states are discrete
is also forbidden by kinematics as long as the $p_i$ are generic.

The only case to consider then is when one + tachyon and one discrete
state appears. The configuration consistent with kinematics
is when the vertices are
$$
\eqalign{
&\VEV{V^-_{j,-m} T^+(-q) T^+(p_{i_1}) \cdots
T^+(p_{i_n}) T^-(k_1) } \qquad
{\rm and}  \cr
&\VEV{V^+_{j,m} T^+(q) T^+(p_{i_{n+1}})\cdots T^+(p_{i_N})}. \cr}
$$
These discrete state correlation functions may be obtained simply
by factorizing the known results for $(N,1)$ tachyon correlators. This
tells us
the first vertex vanishes while the second
is finite.
No additional divergence will arise from the $q$ integration implicit in
\unit\ so this contribution to the imaginary part vanishes.

This proves that particle creation branch cuts are absent in
$(N,1)$ amplitudes at genus one. It is natural to expect this to
be a general feature of $(N,1)$ amplitudes at arbitrary genus.
Unexpectedly the branch cuts observed in the naive effective
tachyon field theory are absent in the full string theory.
This result places strong constraints
on possible string field theories derived from the Liouville
point of view, and provides further evidence for the
equivalence of the $c=1$ matrix model and Liouville theory coupled
to $c=1$ conformal matter.

Finally, consider the $(N,2)$ amplitudes at genus one. In this case
the vertices appearing in the unitarity relations
may include some rather singular discrete state
correlation functions which are not well understood. However,
only considering the intermediate state with two + tachyons
one might expect a finite imaginary part leading to a branch cut.
Let us analyze this case, shown in Fig 3.
Define $n$ to be the number of incoming + tachyons on the left.
Kinematics fixes $k_1=-(n+1)/2$ on-shell, and we also have
$$
q^{\prime} = \sum_{i=1}^n p_i + k_1-q, \qquad\qquad
\E^{\prime} = \sum_{i=1}^n p_i - \E - k_1 -n -1.
\eqn\eq
$$
The discontinuity in the imaginary part of the amplitude coming
from this diagram is
$$
\eqalign{
{\rm Disc} &= {{\pi^{N+1}} \over {n! (N-n)!}} z(p_1)\cdots z(p_N) \cr
& \times\int dq^2~ \delta(\E^2-q^2) \theta(\E) \delta({\E^{\prime}}^2 -
{q^{\prime}}^2) \theta(\E^{\prime}) z(q) z(-q) z(q^{\prime})
z(-q^{\prime}) \cr
&= {{\pi^{N+1}} \over {n! (N-n)!}} z(p_1)\cdots z(p_N)
\int dq ~8 q~{q^{\prime}}^2 \cr
&\times \theta(q) \theta\bigl(\sum_{i=1}^n p_i -q-k_1-n-1\bigr)
\delta\biggl( \bigl(2\sum_{i=1}^n p_i -2q-n-1\bigr)
(2k_1+n+1) \biggr) . \cr }
\eqn\disc
$$
A finite imaginary part might come from integrating out the delta
function. This sets $q^{\prime} = k_1+(n+1)/2$ which vanishes on-shell
causing this contribution to the discontinuity to also vanish. The
presence of the $z$ factors, in particular the decoupling of the
zero momentum mode ($z(0)=0$ ), is crucial here.

Note that there is a contribution to the discontinuity proportional
to $\delta(2k_1+n+1)$
$$
{\rm Disc} = {{2\pi^{N+1}} \over {3~n! (N-n)!}} z(p_1)\cdots z(p_N)
\bigl( \sum_{i=1}^n p_i + k_1\bigr)^3 \delta(2k_1+n+1),
\eqn\eq
$$
which is of the form that would be expected from the resonant limit
of the matrix model prediction for this amplitude, namely
a polynomial in the momenta multiplied by the product of $z$ factors
(the $\delta(2k_1+n+1)$ factor arises in the same way as in the $(N,2)$
amplitudes on the sphere as described above). It will, of course,
be necessary to include the full sum over intermediate states to
see complete agreement between the polynomial terms.

\section{Conclusion}

We have examined some of the constraints imposed by unitarity
on the S-matrix of $c=1$ Liouville theory. These determined
the $(N,2)$ amplitudes on the sphere
in a certain resonant limit and allowed
us to explain the absence of branch cuts in some genus one
amplitudes from the Liouville point of view. Unitarity should
also allow us to derive interesting relations between higher genus
amplitudes,
and may point the way to an off-shell sewing theorem,
but this must await a better understanding of the
role of the discrete states.
\ack
I thank A.M. Polyakov and I.R. Klebanov for many useful discussions
and suggestions.
This research was supported in part by DOE grant DE-AC02-76WRO3072,
NSF grant PHY-9157482, and James S. McDonnell Foundation grant No.
91-48.

\refout
\bye